\documentclass{jpsj2}

\usepackage{graphicx}

\def\runauthor{Yunori  {\sc Nisikawa}, 
Manabu {\sc Usuda}, Jun-ichi {\sc Igarashi}, 
Hironobu {\sc Shoji}, and Toshiaki {\sc Iwazumi}}
\title
{Resonant Inelastic X-Ray Scattering at the K Edge of Ge
 }
\author
{ 
Yunori  {\sc Nisikawa}\footnote{E-mail
address:nisikawa@spring8.or.jp}$^1$, Manabu {\sc
Usuda}$^1$, Jun-ichi {\sc Igarashi}$^1$,\\
Hironobu {\sc Shoji}\footnote{Present address: Semiconductor Energy Laboratory Co., Ltd., 398 Hase, Atsugi, Kanagawa 243-0036.}$^2$,    
and Toshiaki {\sc Iwazumi}$^3$
}
\inst
{$^1$Synchrotron Radiation Research Center, Japan Atomic Energy
Research Institute, Mikazuki, Sayo, Hyogo 679-5148\\
$^2$Institute of Industrial Science, University of Tokyo, Tokyo 153-8505\\
$^3$Photon Factory, Institute of Materials Structure Science, 1-1 Oho, Tsukuba, Ibaraki 305-0801}

\recdate
{
\today
}
\abst{We study the resonant inelastic x-ray scattering (RIXS) at the $K$ edge
of Ge. We measure RIXS spectra with systematically varying
momenta in the final state.
The spectra are a measure of exciting an electron-hole pair.
We find a single-peak structure (except the elastic peak)
as a function of photon energy,
which is nearly independent of final-state momenta.
We analyze the experimental data by means of the band structure calculation.
The calculation reproduces well the experimental shape,
clarifying the implication of the spectral shape.}
\kword{resonant inelastic X-ray scattering, hard X-ray, Ge}

\begin{document}
\sloppy
\maketitle
\section{Introduction}
Inelastic x-ray scattering is a promising method 
to study electronic structures in matters.
Since intensities are much smaller than those on the elastic part,
high-brilliance synchrotron sources are necessary.
It is advantageous to use a resonant enhancement 
by tuning photon energy near absorption edge.
This resonant inelastic x-ray scattering (RIXS)
is described by a second-order optical process,
in which a core electron is excited by an incident photon and then
this excited state decays by emitting a photon to fill the core hole.
Thereby electron-hole pairs remain in the final state.
We can cover the momentum of the final state over the wide range
by using the K edge of transition metals.
Such RIXS experiments have recently been carried out 
on cuprates\cite{HASAN,KIM} and manganites\cite{INAMI},
having revealed characteristics of charge excitations through
the momentum dependence of spectra. 

On semiconductors and insulators such as silicon\cite{MA}, 
graphite\cite{CARLISLE}, boron nitride\cite{JIA}, and others\cite{KOTANI},
RIXS experiments have already been performed,
where corresponding photon energies are in the soft x-ray region.
The spectra are a measure of exciting an electron-hole pair.
Band structure calculations have reproduced well the spectral shape
due to small electron correlations.
In this paper, we carry out a RIXS experiment using the K edge in Ge.
Since corresponding photon energies are in the hard x-ray region,
we expect that a new insight would come out through the momentum dependence
of the spectra.
A similar experiment has already been reported\cite{KAPROLAT}.
The present experiment is much thorougher with systematically varying 
momenta in the wide range. 
Against our expectation, the experimental spectra as a function 
of photon energy show a broad single peak (except the elastic peak), 
and keep almost the same shape with varying momenta.

We analyze the experimental data by calculating the spectra 
within the local density approximation (LDA).
We disregard the $1s$ core-hole potential in the intermediate state,
expecting that this neglect would cause only a minor error\cite{Veenendaal}. 
This is partly justified from our calculation of the absorption coefficient,
which agrees near the edge with the experimental one, as shown below.
The final state consists of one electron in the conduction band and
one hole in the valence band. If the core-hole level in the
intermediate state is sharp enough, we can select the momentum 
of the excited electron in the conduction band by sharply 
tuning the incident-photon energy,
because the denominator in the formula of the second order process
(eq.~(\ref{ds})) selects a particular process through enhancement. 
Then, selecting the final-state momentum by setting 
the scattering geometry, we can specify the momentum of the hole 
in the valence band. In this situation, one expects sharp peaks
at the energy of the electron-hole pair, as a function of photon energy.
On the other hand, if the core-hole level is broad,
many channels are opened to excite an electron 
with different momenta in the conduction band.
We need to sum up each contribution.
We calculate the RIXS spectra with varying values of the core-level width.
This leads to an examination of determining the band structure from RIXS
spectra\cite{rf:MaTheory}.
The $1s$ core level of Ge is rather broad with width of about 2 eV.
Summing up each contribution of band-to-band transitions, 
we obtain the spectra in good agreement with the experiment.
Thus the analysis based on the band structure calculation demonstrates
the origin of the spectra.

This paper is organized as follows. 
Experimental details are described in \S \ref{exp},
and the formalism for calculating the spectra is described in \S \ref{forma}.
In \S \ref{compe}, the experimental data are presented and compared
with the calculated results. Section 5 is devoted 
to concluding remarks.

\section{Experimental Measurements}\label{exp}
\begin{figure}
\includegraphics[width=15cm,height=19cm]{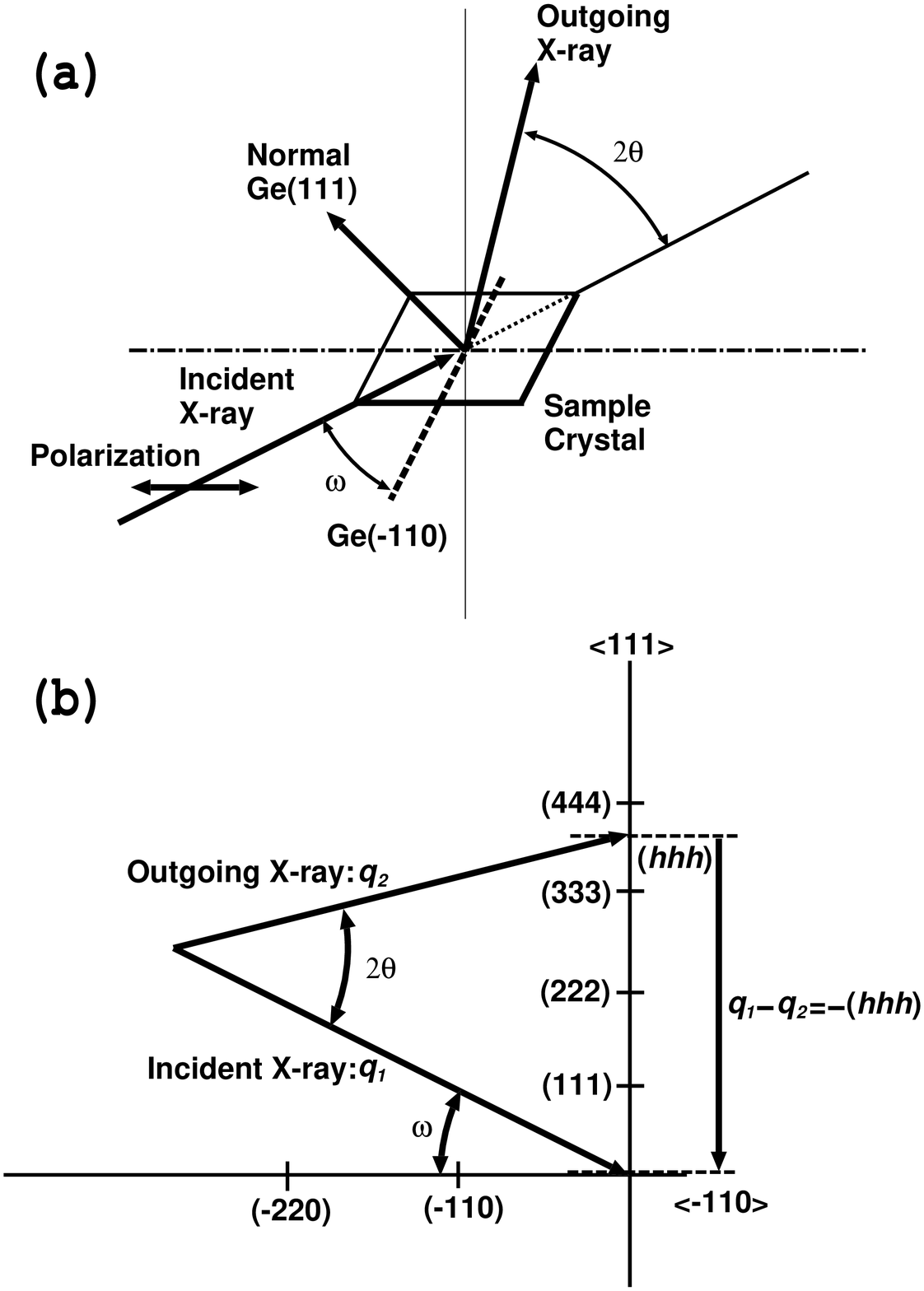}
\caption{(a)Experimental set up for scattering. 
(b)Geometrical relations between ${\bf q}_{1}, {\bf q}_{2}$ and crystal axis of Ge. }
\label{fig:exp}
\end{figure}

The experimental set up is shown in Fig.~\ref{fig:exp} (a).
The geometrical arrangement ($\omega,\theta$) is determined as shown in
Fig.~\ref{fig:exp}(b), for the momentum transfer from
the incident X-ray with momentum ${\bf q}_1$ to the outgoing X-ray with
momentum ${\bf q}_2$. Experiment was performed at the bending magnet 
beam-line (7C) of the Photon Factory, Institute of Materials 
Structure Science.
This beamline consists of a Si (111) 
double-crystal monochromator with the sagittal focusing 
mechanism and a focusing double mirror, and provides 
linear polarized x-rays.
The incident beam size at the sample position was about
1.5 mm in horizontal and 1 mm in vertical;
the photon flux was estimated to be 
$\sim 1 \times$ 10$^{11}$ photons/s.
The dimension of the Ge single crystal was 
10 mm $\times$ 10 mm $\times$ 5 mm.
Emitted x-rays were analyzed by 
means of a cylindrically-bent InSb(880) crystal and measured 
by a position sensitive proportional counter.
The total energy resolution of the emission 
spectrum in the present apparatus was 3.0 eV.
More details of the spectrometer has already 
been described in ref.~\ref{expspec}.

\section{Formulation}\label{forma}
\subsection{Double Differential Scattering Cross-Section}
RIXS is described by a second-order process.
Using the generalized Fermi's golden rule where the interaction between 
light and electrons is treated by second order perturbation theory,
we obtain the double differential scattering cross-section.
%
%
\begin{equation}\label{ds}
\frac{d^{2}\sigma}{d\omega_{2} d\Omega_{2}}\propto
\sum_{f}\left|
\sum_{m}
\frac{<f|T^{\prime}|m><m|T|i>}{E_{m}-E_{i}-\hbar\omega_{1}-i\Gamma_{m}/2}
\right|^{2}
\delta(E_{f}-E_{i}-\hbar\omega_{1}+\hbar\omega_{2}).
\end{equation}
Here
$T\equiv {\bf e}_{1}\cdot\sum_{j}{\bf p}_{j}e^{i{\bf q}_{1}\cdot
{\bf r}_{j}}$, $T^{\prime}\equiv {\bf e}_{2}\cdot\sum_{j}{\bf p}_{j}e^{-i{\bf q}_{2}\cdot {\bf r}_{j}}$
and  ${\bf e}_{i}$, $\hbar\omega_{i}$ and $\hbar {\bf q}_{i}$ denote
the polarization vector, the energy and the momentum vector
of the incident ($i=1$) and emitted ($i=2$) photon, respectively, and
${\bf r}_{j}, {\bf p}_{j}$ 
denote the position and momentum of electron $j$. 
Kets $|i>$, $|m>$ and $|f>$ represent initial, intermediate and finale
states of the material system, respectively, with their energies
$E_{i}$, $E_{m}$, and $E_{f}$.
The $\hbar/\Gamma_{m}$ is the life time of the intermediate state $|m>$.

For germanium, an independent particle treatment for electron system seems to 
be appropriate, since electron correlations are weak~\cite{rf:MaTheory}.
For usual situations, the initial state $|i>$ is assumed to be
the ground state $|\Omega>$.
The incident photon with momentum ${\bf q}_1$ excites an electron 
from the $1s$ core state $c$ with crystal momentum ${\bf k}_c$
to the conduction band (whose band index is $e$) 
with crystal momentum ${\bf k}(={\bf q}_1+{\bf k}_c)$.
Neglecting the core-hole potential in the intermediate state, we express
the intermediate state 
by $|m>=a_{{\bf k},e}^{\dagger}a_{{\bf k}_c,c}|\Omega>$,
where $a_{\xi}^{\dagger}(a_{\xi})$ is the creation(annihilation) operator
for the electron of $\xi$ state.
Then, the intermediate state decays with emitting photon with momentum 
${\bf q}_2$ and annihilating an electron in the valence band 
(whose band index is $h$) with crystal momentum 
${\bf k}^{\prime}(={\bf q}_2+{\bf k}_c)$ to fill the core hole. 
Therefore, we have the final state 
$|f>=a_{{\bf k},e}^{\dagger}a_{{\bf k}^{\prime},h}|\Omega>$.
Since the final state uniquely determines the intermediate state, 
eq.~(\ref{ds}) is rewritten as 
\begin{equation}\label{dsS}
\frac{d^{2}\sigma}{d\omega_{2} d\Omega_{2}}\propto
\sum_{({\bf k},e),({\bf k}^{\prime},h)}
\frac
{\left|\sum_{a}\exp(i({\bf q}_{1}-{\bf q}_{2}){\bf R}_{a})\overline{t_{a}({\bf k}^{\prime},h|{\bf e}_{2})}
t_{a}({\bf k},e|{\bf e}_{1})\right|^{2}}
{(\epsilon_{e}({\bf k})-\epsilon_{c}-\hbar\omega_{1})^{2}+\Gamma_{m}^{2}/4}
\delta^{{\bf G}}_{{\bf \Omega},{\bf k}-{\bf k}^{\prime}}
\delta(\epsilon_{e}({\bf k})-\epsilon_{h}({\bf k}^{\prime})-\hbar\omega),
\end{equation}
where $\hbar\omega=\hbar\omega_{1}-\hbar\omega_{2}$, and ${\bf
 \Omega}={\bf  q}_{1}-{\bf q}_{2}$ are energy and momentum of the
 final state. $\epsilon_{e}({\bf k})$, $\epsilon_{h}({\bf k}^{\prime})$ and
 $\epsilon_{c}$
are the energy of the excited electron with crystal momentum ${\bf k}$
in the conduction band, that of the hole with crystal momentum 
${\bf k}^{\prime}$ in the valence band, and that of the core state, 
respectively.
The crystal momentum conservation for the whole process is contained
in the factor of Kronecker $\delta$,
\begin{equation}
\delta^{\bf G}_{{\bf \Omega},{\bf k}-{\bf k}^{\prime}}
\equiv  \left\{
\begin{array}{@{\,}ll}
0 & \mbox{: ${\bf \Omega}-({\bf k}-{\bf k}^{\prime})\notin {\bf G}$ }\\
1 & \mbox{: ${\bf \Omega}-({\bf k}-{\bf k}^{\prime})\in {\bf G}$ }
\end{array}
\right.,
\end{equation}
where ${\bf G}$ is the set of reciprocal lattice vectors. 
Overlined quantities indicate their complex conjugates.
Within the dipole approximation, $t_{a}({\bf p},\tau|{\bf e})=
\int d{\bf r}\overline{\psi_{{\bf p},\tau}({\bf r})}{\bf e}\cdot \hat{{\bf
p}}\phi_{a}^{1s}({\bf r}-{\bf R}_{a})$, 
where ${\bf R}_{a}$, $\psi_{{\bf p},\tau}$
and $\phi_{a}^{1s}$ are the position vector of atom $a$ in unit cell, 
Bloch-wave function of a electron in the $\tau$-band with 
crystal momentum ${\bf p}$, and 1$s$-atomic orbital, respectively.
These quantities are evaluated on the basis of the band structure calculation
described in the following sub-section.

\subsection{Band Structure Calculation of Ge}
\begin{figure}
\includegraphics[width=10cm,height=10cm]{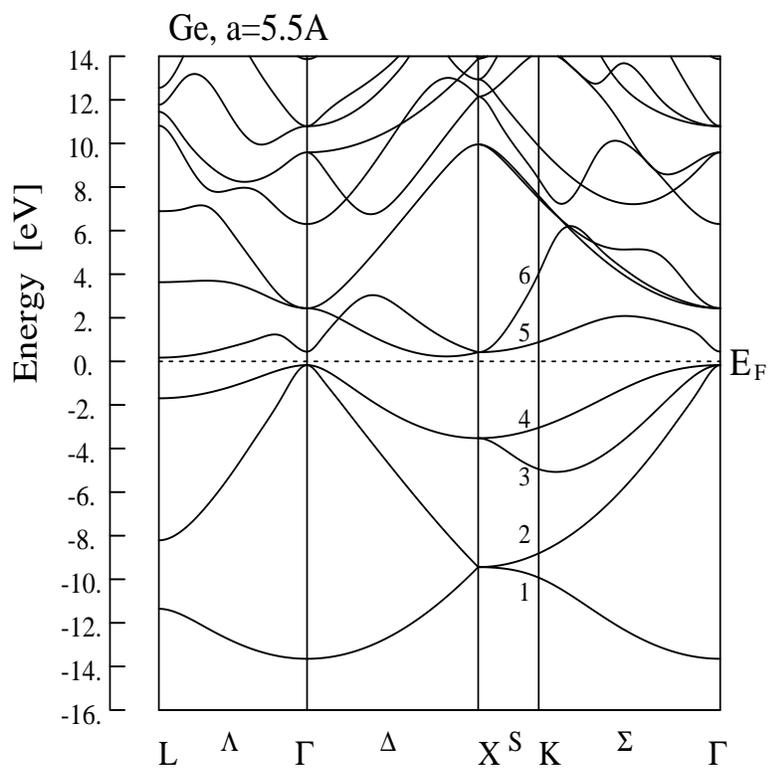}
\caption{Energy vs. momentum relation of Ge calculated by the FLAPW method 
in the LDA scheme.}
\label{fig:band}
\end{figure}

We perform a band structure calculation using
the full-potential linearized augmented-plane-wave (FLAPW) method
\cite{JF84} within the LDA.
 The local exchange-correlation functional of Vosko, Wilk and Nusair
 is employed~\cite{VWN80}. The angular momentum in the spherical-wave expansion 
 is truncated at
 $l_{\rm max}=6$ and $7$ for the potential and wave function, respectively.
 The energy cutoff of the plane wave is 12 Ry for the wave function.
It is well known that the density-functional theory underestimates
 bandgaps. For Ge, the density-fuctional calculation
 fails to obtain a finite bandgap: the valence-band top and
 the conduction-band bottom overlap in energy at the $\Gamma$ point,
 leading to a negative value of bandgap. Experiments suggests a finite bandgap,
 0.744 eV.
 Many-body theories based on the Green function such as
 the $GW$ calculation 
may be necessary to reproduce
 a collect size of the band gap. For the present purpose, however,
 it is sufficient to employ a simple remedy for the problem: 
 carrying out a calculation on a system of a smaller lattice constant. 
 The bandgap becomes larger for smaller lattice constants, since
 the splitting between bonding and antibonding bands becomes larger.
 We obtain a bandgap of 0.35 eV between the $\Gamma$ and $X$ points 
 for the lattice constant $a=5.5\AA$ ($a=5.658\AA$ in the real system). 
 Figure \ref{fig:band} shows the energy vs. momentum relation
 thus evaluated. The energy band is labeled by attached numbers for later use.
 Note that a slight change of the lattice constant
 has little influence on the dispersion relation except the band gap.

\section{Comparison of experimental results and theoretical calculations}\label{compe}
We present the experimental data in comparison with the calculated spectra.
The calculation is made by following the procedure mentioned above
with assuming
$\Gamma_{m}=\Gamma=2.0$eV (for all $m$) and $\epsilon_{c}=-11,103$eV 
(measured from Fermi level).

\subsection{Absorption coefficient at the K edge}
\begin{figure}
\includegraphics[width=15cm,height=13cm]{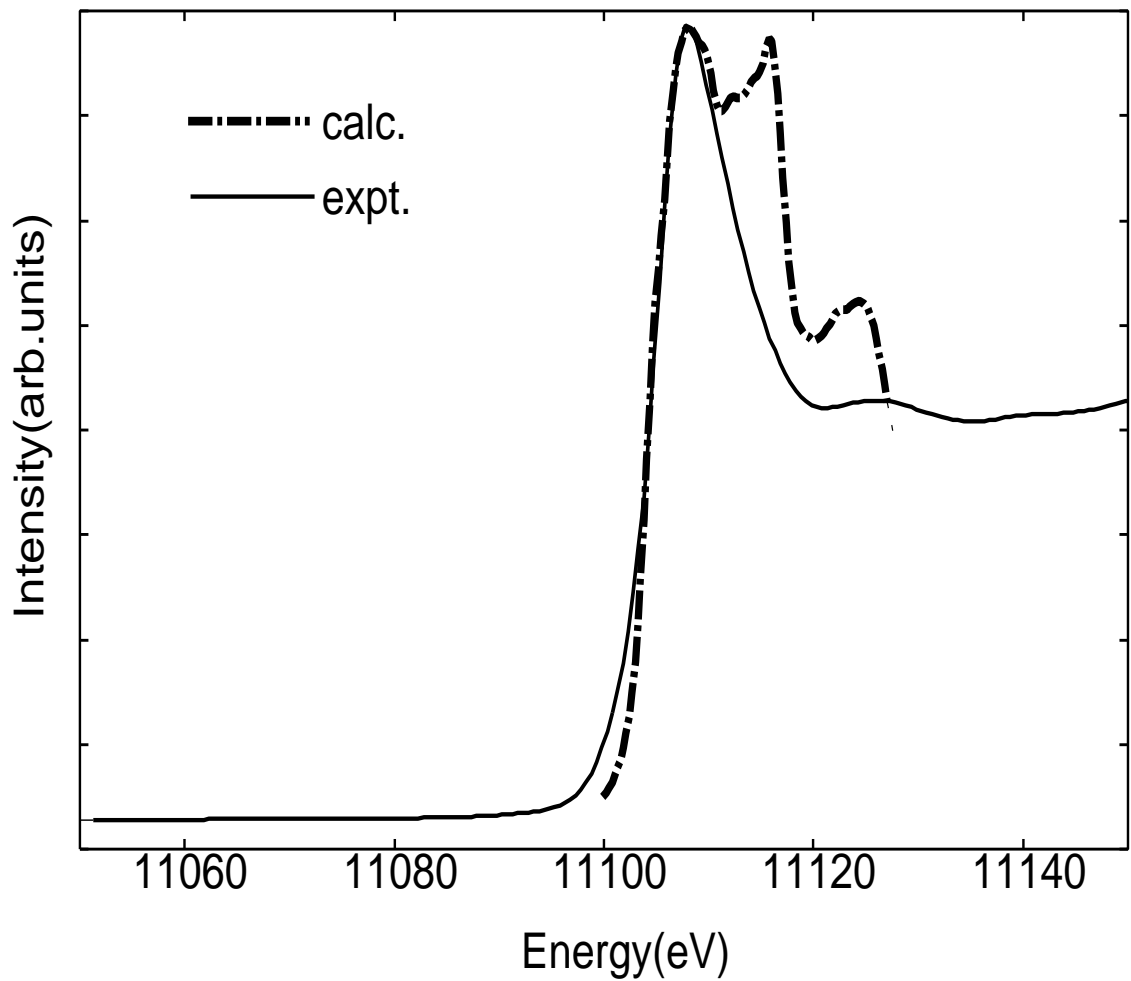}
\caption{Absorption coefficient calculated at the K edge in comparison with
the experiment (Ref.\ref{kokubu}).}
\label{fig:GeK}
\end{figure}

Before going to the RIXS spectra, we first calculate the absorption
coefficient at the K edge. Figure \ref{fig:GeK} shows the calculated result
in comparison with experimental results 
by Kokubun {\it et al.}~\cite{rf:GeKabs1,rf:GeKabs2}.
The experiment shows no sign of core exciton.
The calculated curve is in agreement with the experimental one in the energy 
region between 11,100 eV and 11,110 eV,
indicating that the effect of core hole potential seems unimportant
in the energy region of our interest.
The calculation predicts extra peaks at about 11,115 eV and 11,125 eV,
deviating from the experimental curve.
This fault come from a linearizing procedure in the FLAPW method, 
which becomes less accurate for high energy states.

\subsection{RIXS spectra}
\begin{figure}
\includegraphics[width=15cm,height=19cm]{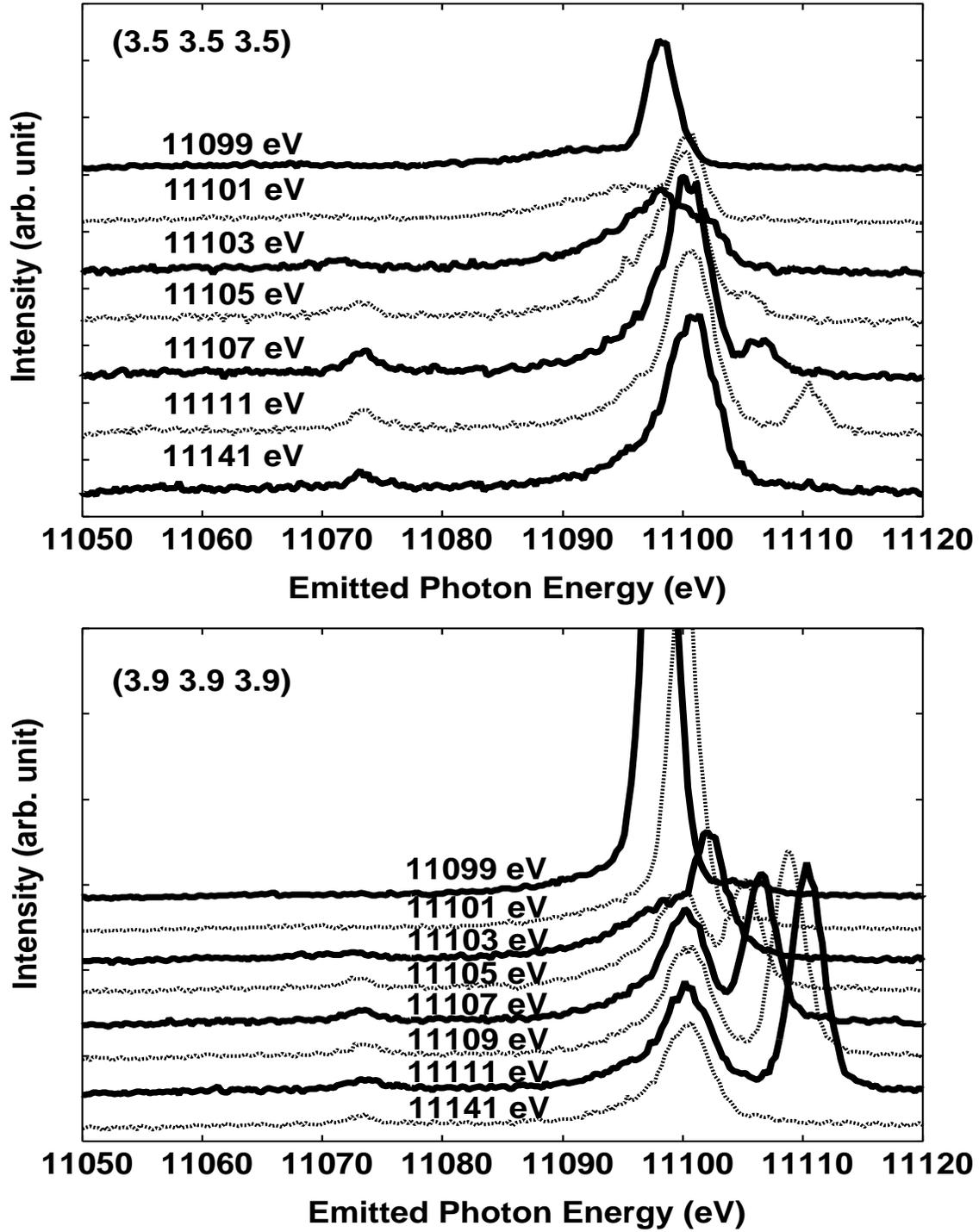}
\caption{Experimental RIXS spectra as a function of emitted photon energy
$\hbar\omega_2$ at (a) final-state momentum ${\bf \Omega}=-(3.5,3.5,3.5)$
and (b) $-(3.9,3.9,3.9)$. The incident photon energy $\hbar\omega_1$ 
varies from $11,099$ eV to $11,141$ eV.}
\label{fig:expspec}
\end{figure}

\begin{figure}
\includegraphics[width=15cm,height=19cm]{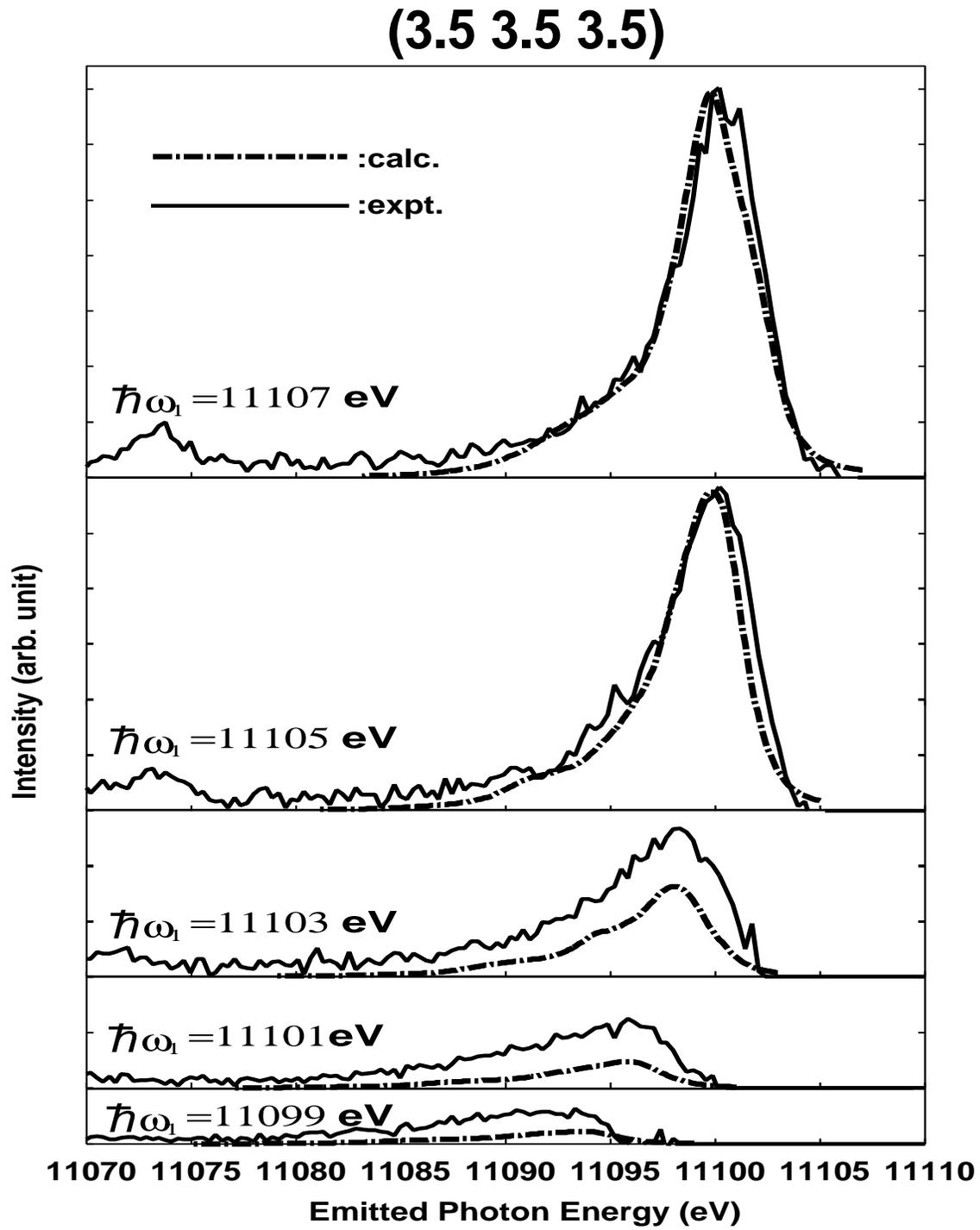}
\caption{Inelastic contribution of RIXS spectra as a function of
emitted photon energy $\hbar\omega_2$.
Incident photon energy $\hbar\omega_1$ varies from $11,099$ eV to $11,107$ eV.
}
\label{fig:energy}
\end{figure}

Figure \ref{fig:expspec} shows experimental RIXS spectra with varying 
incident photon energy between 11,099 eV and 11,141 eV. 
The final-state momenta are set to be
${\bf\Omega}=-(3.5,3.5,3.5)$ and $-(3.9,3.9,3.9)$.
There are two peaks.
One is an elastic peak, which is larger for the 
final-state momentum close to a reciprocal lattice vector.
Another is an inelastic peak.
Both get closer for the incident photon energy getting close to 
the threshold. 
Since the tail of elastic contribution masks the inelastic contribution,
we have to subtract the former from experimental data in order to extract
the latter contribution. 
Assuming that the elastic contribution dominates the intensity
in the region of $\hbar\omega=\hbar\omega_1-\hbar\omega_2 < 0$,
we evaluate the elastic contribution for $\hbar\omega > 0$
by the value at $-\hbar\omega$.
This may be a slight overestimate of the elastic contribution in the 
region of $\hbar\omega > 0$.
Figure \ref{fig:energy} shows the inelastic contribution 
evaluated this way in comparison with the calculation.
The calculated intensity is normalized such that the peak height 
coincides with the experimental one at $\hbar\omega_1=11,107$ eV.
We find a single broad peak, whose intensity increases with increasing
incident photon energies.
The calculation reproduces well such experimental behavior.
This behavior of the peak intensity corresponds to 
the behavior of absorption coefficient at the K edge 
in the energy region between 11,100 eV and 11,107 eV where 
absorption coefficient increases with increasing
incident photon energies (See Figure \ref{fig:GeK}).

\begin{figure}
\includegraphics[width=17cm,height=18cm]{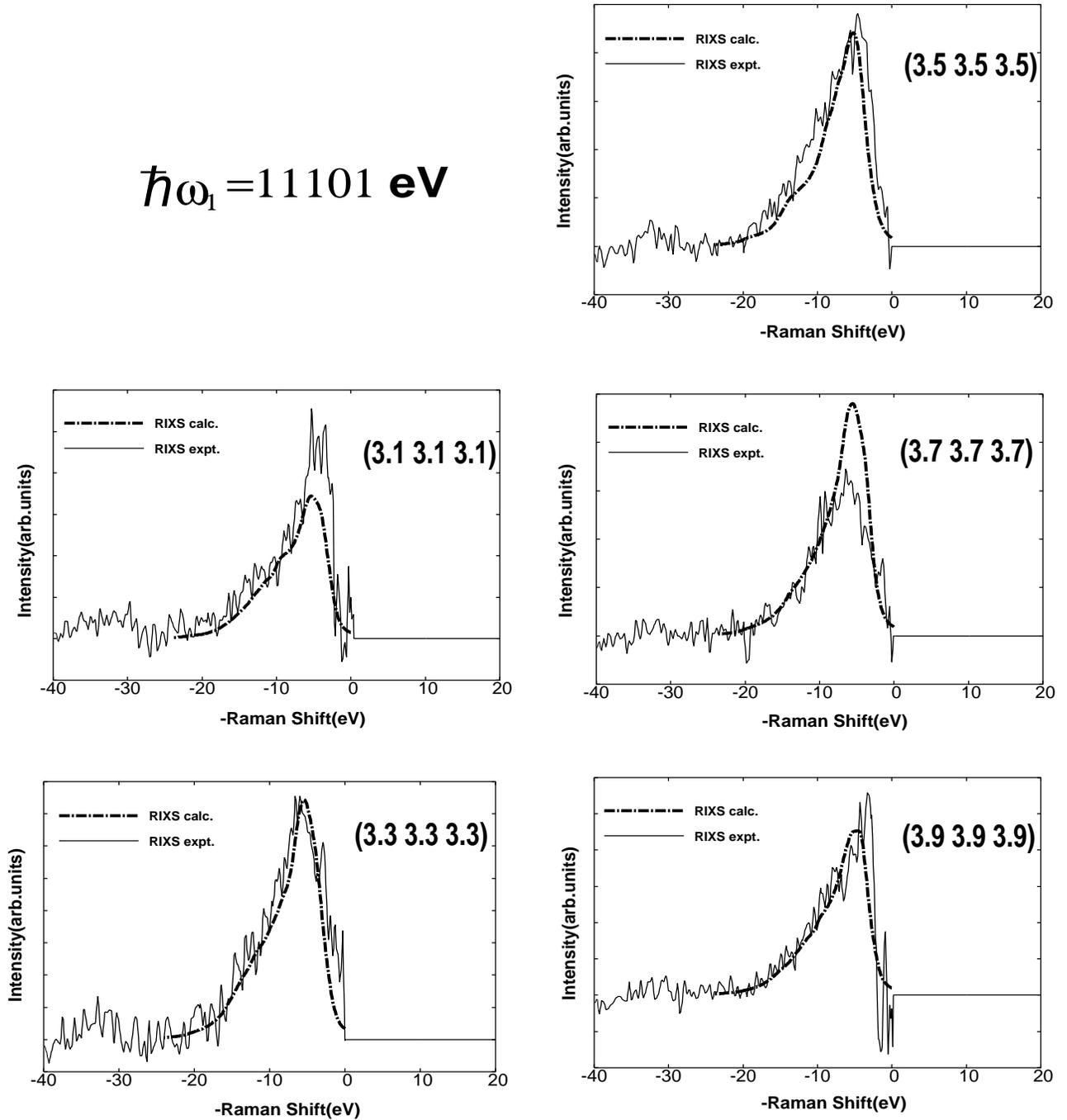}
\caption{Inelastic contribution of RXS spectra as a function of Raman shift
$\hbar\omega=\hbar\omega_1-\hbar\omega_2$ with varying final-state momenta.
}
\label{fig:moment}
\end{figure}

Figure \ref{fig:moment} shows RIXS spectra as a function of the Raman shift 
($=\hbar\omega$) for $\hbar\omega_1=11,101$ eV.
The final-state momenta vary from ${\bf\Omega}=-(3.1,3.1,3.1)$ 
to $-(3.9,3.9,3.9)$. The elastic contribution is subtracted in the same way
as above.
The calculated results are in good agreement with the experimental ones.
The spectra are nearly independent of final-state momenta.
The reason will be discussed in the following subsection.

\begin{figure}
\includegraphics[width=8cm,height=19cm]{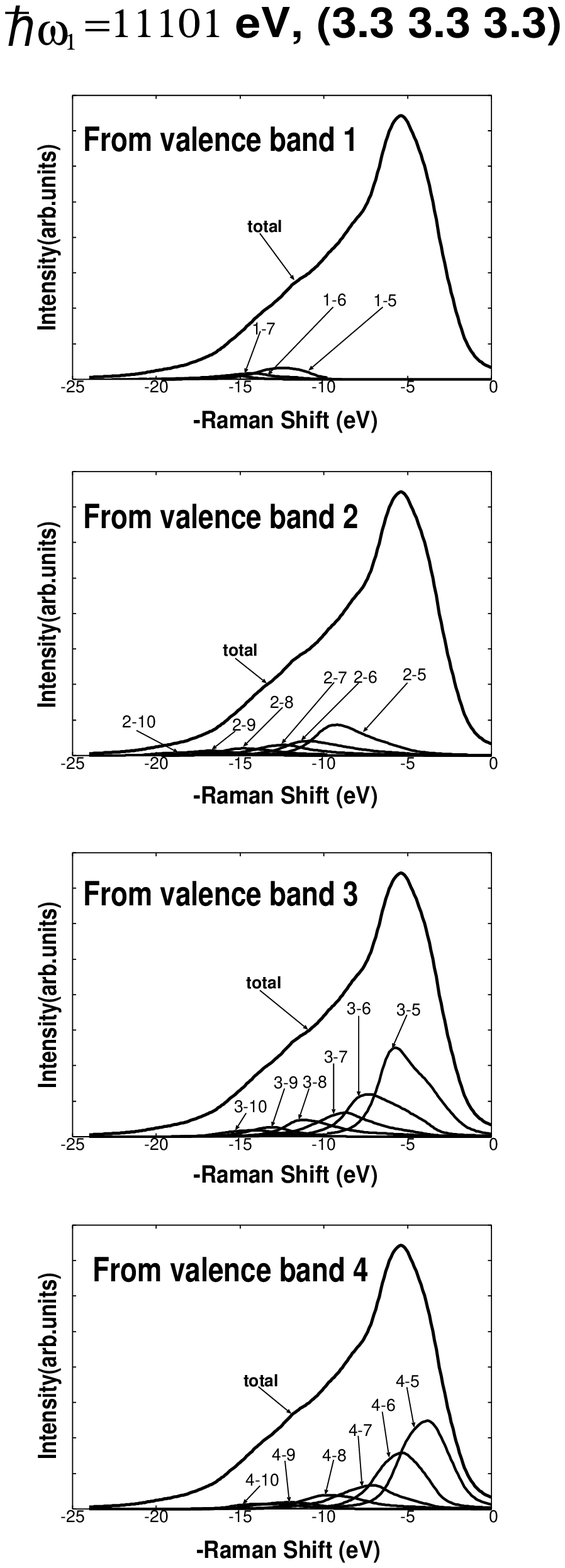}
\caption{Decomposition of RIXS spectra into contributions of band-to-band
excitations, for $\hbar\omega_1=11,101$ eV, ${\bf  \Omega}=-(3.3,3.3,3.3)$.
Attached number $m-n$ represents a transition from band $m$ to band $n$.
}
\label{fig:decomp}
\end{figure}

The spectra are decomposed into each contribution of band-to-band 
transitions specified by band indices.
Figure \ref{fig:decomp} shows the result for ${\bf\Omega}=-(3.3,3.3,3.3)$, 
$\hbar\omega_1=11,101$ eV.
We notice that the peak is mainly composed of transitions 
from the bands 3 and 4 to the band 5.

\subsection{Reconsideration of the idea of band-structure determination
from RIXS spectra}\label{discuss}
We examine more closely eq.~(\ref{dsS}) to clarify the origin of
the spectral shape.
This leads to the idea of band-structure 
determination from RIXS spectra proposed by Ma~\cite{rf:MaTheory}.

\begin{figure}
\includegraphics[width=15cm,height=13cm]{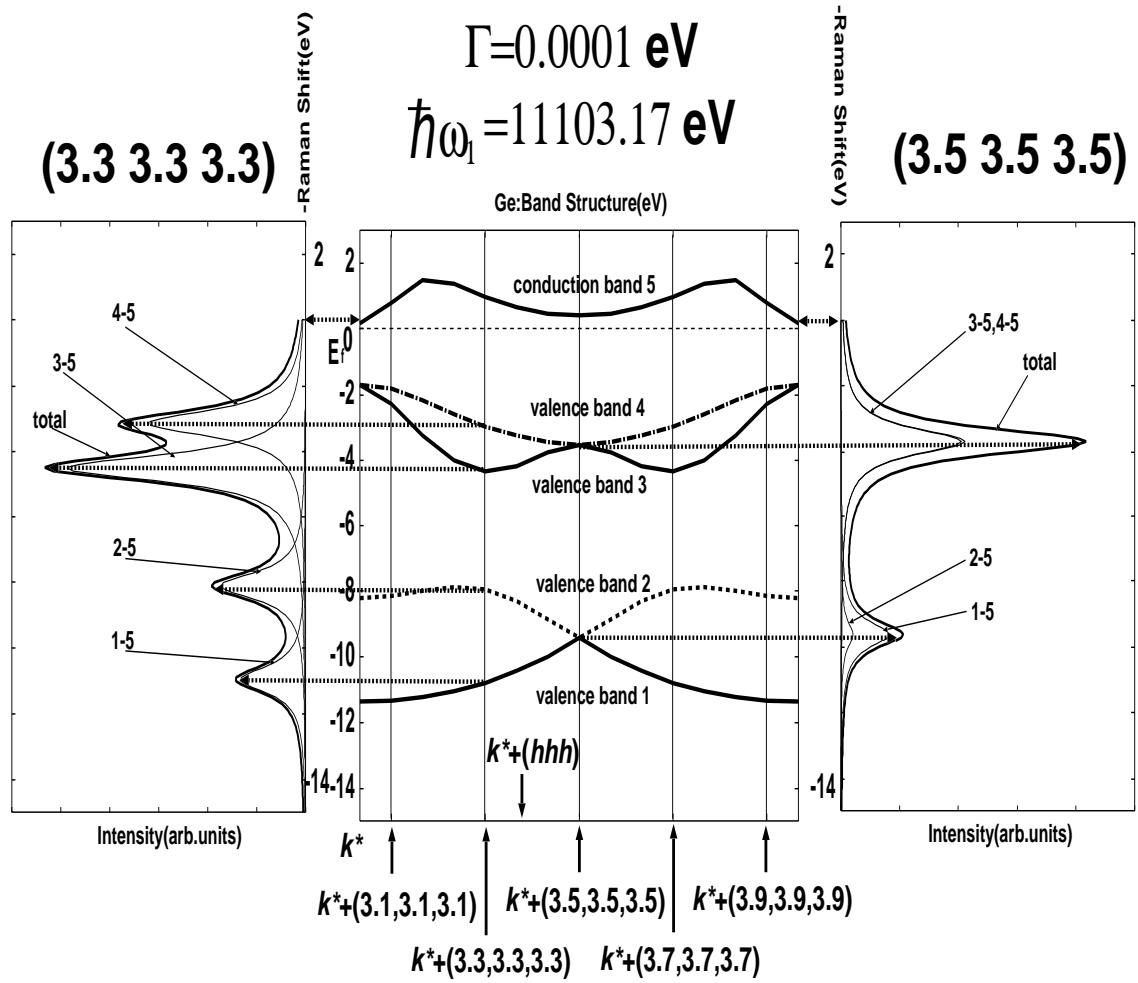}
\caption{RIXS spectra calculated by assuming $\Gamma=0.0001$ eV.}
\label{fig:map}
\end{figure}

We start by defining a function (here we neglect $a$-dependence of 
$t_{a}({\bf p},\tau|{\bf e})$ and write $t({\bf p},\tau|{\bf e})$),
\begin{equation}
\label{eq:phi}
\psi({\bf k},e)\equiv\frac
{\left|t({\bf k},e|{\bf e}_{1})\right|^{2}}
{(\epsilon_{e}({\bf k})-\epsilon_{c}-\hbar\omega_{1})^{2}+\Gamma^{2}/4}.
\end{equation}
Since the core hole can carry momentum, the excited electron can take any
value of momentum according to ${\bf k}={\bf q}_1+{\bf k}_c$.
Here, we consider the following limit case 
opposite to the real situation in Ge.
If the value of $\Gamma$ is much smaller than
the band-with of a branch $e^{*}$ in the conduction band,
the denominator in eq.~(\ref{eq:phi}) 
can become very small for a certain value ${\bf k}^*$ of ${\bf k}$. 
In the case of band structure of Ge, 
the L point is selected when the incident photon energy is 
tuned at the absorption edge.
Thus eq.~(\ref{eq:phi}) is approximated by
\begin{equation}\label{eq:psidelta}
\psi({\bf k},e)\propto \delta({\bf k}-{\bf k}^{*})\delta_{e,e^{*}}.
\end{equation}
Substitution of this relation into eq.~(\ref{dsS}) leads to
\begin{equation}\label{eq:map}
\frac{d^{2}\sigma}{d\omega_{2} d\Omega_{2}}\propto
\sum_{h}
\left|t({\bf k}^{*}-{\bf\Omega},h|{\bf e}_{2})
\right|^{2}\delta(\epsilon_{e^{*}}({\bf k}^{*})-\epsilon_{h}({\bf k}^{*}
-{\bf\Omega})-\hbar\omega).
\end{equation}
The spectra consist of several $\delta$-functions 
as a function of $\omega$, whose positions
move with varying values of ${\bf\Omega}$.
Figure \ref{fig:map} illustrates how the band is determined when 
the incident photon energy is tuned at the absorption edge ( ${\bf
k}^{*}$ is at L-point).
Thereby we can trace the valence band structure from the peak position
by means of scanning ${\bf\Omega}$.

Next we consider the following limit case opposite to discussion
mentioned above.  
For germanium, the value of $\Gamma$ at the K-edge is estimated about 2 eV. 
This value is nearly equal to 
the band-with of the lowest conduction band of Ge.
In such a situation, one may safely put the ${\bf k}$-dependence 
on $\psi({\bf k},e)$, that is, put
\begin{equation}\label{psiconst}
\psi({\bf k},e)\simeq \psi(e).
\end{equation}
This relation leads to an approximate expression to
the differential cross-section,
\begin{equation}\label{eq:gecase}
\frac{d^{2}\sigma}{d\omega_{2} d\Omega_{2}}\propto
\sum_{e}\psi(e)F(\overline{\epsilon_{e}}-\hbar\omega),
\end{equation}
with
\begin{equation}
  F(\hbar\omega)\equiv\sum_{({\bf k}^{\prime},h)}
 \left|t({\bf k}^{\prime},h|{\bf e}_{2})\right|^{2}\delta(\hbar\omega-\epsilon_{h}({\bf k}^{\prime})), 
\end{equation}
where $\overline{\epsilon_{e}}$ is the average of $\epsilon_{e}({\bf k})$ 
over ${\bf k}$ in the Brilloin zone.
Since $|t({\bf k}^{\prime},h|{\bf e}_{2})|^{2}$
selects $p$ symmetric states within the dipole approximation,
$F(\hbar\omega)$ is analogous to the density of states (DOS) projected onto 
$p$-symmetric states.
The spectra are given by a superposition of the $p$-DOS,
and this naturally explains why the spectra are nearly independent of
final-state momenta.

\begin{figure}
\includegraphics[width=15cm,height=15cm]{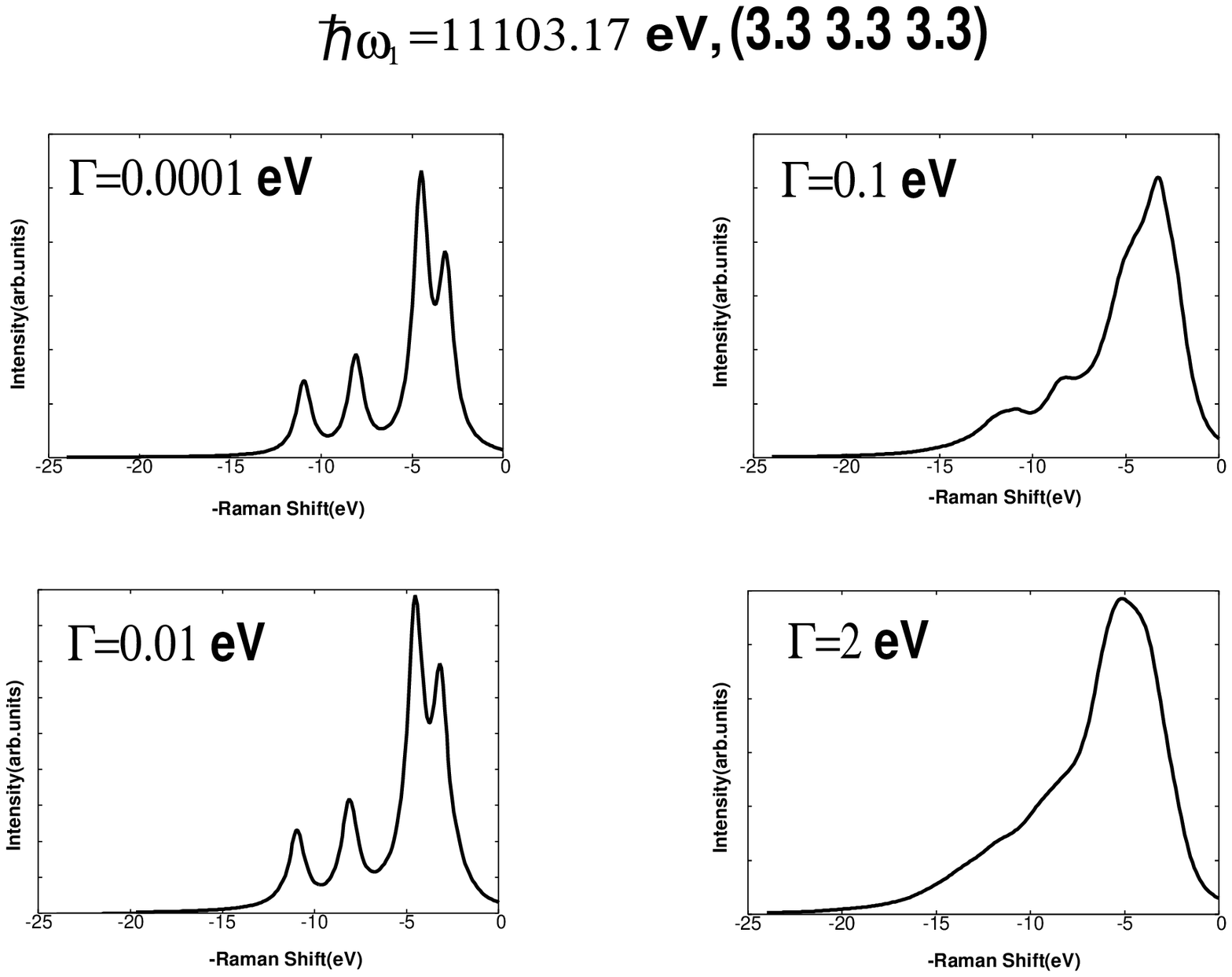}
\caption{$\Gamma$ dependence of RIXS spectra}
\label{fig:Gamma}
\end{figure}

In Figure.~\ref{fig:Gamma}, we show $\Gamma$-dependence of RIXS
spectra. From above discussion, 
the spectra for $\Gamma=0.0001, 0.01, (0.1)$ eV should be interpreted 
according to eq.(~\ref{eq:map}) and the spectrum for $\Gamma=2$ eV should be 
interpreted according to eq.(~\ref{eq:gecase}).


  
\section{Concluding Remarks}\label{conrema}
We have carried out a RIXS experiment at the $K$ edge of Ge
with systematically varying final-state momenta.
We have obtained a broad inelastic peak as a function of photon energy,
whose shape is nearly independent of final-state momenta.
We have analyzed the experimental data on the basis of the band structure 
calculation, summing up each contribution of band-to-band transitions.
The band calculation has reproduced well the experimental data,
demonstrating the origin of the spectra.

We have examined a possibility of determining band structure
from RIXS experiment.
For smaller values of core-level widths, we can trace 
the energy dispersion with varying final-state momenta.
To obtain sizable momentum change, however,
photon energy is required 
to be larger than 2-3 keV. But, since corresponding core-level 
widths are usually large, the spectra are likely to be 
independent of final state momenta, 
thus making it difficult to determine the band structure.

\section*{Acknowledgments}
Some of the author (Y.N, M.U and J.I) 
greatly thank to Prof. N. Hamada for allowing us to use his
FLAPW code.

\newpage

\begin{thebibliography}{99}

\bibitem{HASAN}
        M. Z. Hasan, E. D. Isaacs, Z.-X. Shen, L. L. Miller,
        K. Tsutsui, T. Tohyama, and S. Maekawa,
        Science {\bf 288} (2000) 1811.
\bibitem{KIM}
        Y. J. Kim, J. P. Hill, C. A. Burns, S. Wakimoto, R. J. Birgeneau,
        D. Casa, T. Gog, and C. T. Venkataraman,
        Phys. Rev. Lett. {\ 89} (2002), 177003.
\bibitem{INAMI}
        T. Inami, T. Fukuda, J. Mizuki, S. Ishihara, H. Kondo,
        H. Nakao, T. Matsumura, K. Hirota, Y. Murakami,
        S. Maekawa, and Y. Endoh,
        Phys. Rev. B {\bf 67} (2003) 045108.
\bibitem{MA}
        Y. Ma, K. E. Miyano, P. L. Cowan, Y. Aglitzkiy, and B. A. Karlin,
        Phys. Rev. Lett. {\bf 74} (1995) 478.
\bibitem{CARLISLE}\label{CARLISLE}
        J. A. Carlisle, E. L. Shirley, E. A. Hudson, L. J. Terminello,
        T. A. Calcott, J. J. Jia, D. L. Ederer, R. C. C. Perera,
        and F. Himpsel, Phys. Rev. Lett. {\bf 74} (1995) 1234.
\bibitem{JIA}
        J. J. Jia, T. A. Callcott, E. L. Shirley, J. A. Carlisle,
        L. J. Terminello, A. Asfaw, D. L. Ederer, F. J. Himpsel,
        and R. C. C. Perera, Phys. Rev. Lett. {\bf 76} (1996) 4054.
\bibitem{KOTANI}
        A. Kotani and S. Shin, Rev. Mod. Phys. {\bf 73} (2001) 203.
\bibitem{KAPROLAT}
        A. Kaprolat and W. Sch\"ulke, Applied Phys. A {\bf 65} (1997) 169.
\bibitem{Veenendaal}
At the $K$ edge of graphite, a core exciton is observed.
The effect on the RIXS spectra is studied by Carlisle {\it et al.}(Ref.~\ref{CARLISLE}) and by
        M. van Veenendaal and P. Carra,
        Phys. Rev. Lett. {\bf 78} (1997) 2839.
\bibitem{rf:MaTheory} Y. Ma, Phys. Rev. B {\bf 49} (1993) 5799
\bibitem{rf:expspec}
T, Iwazumi, K. Kobayashi, S. Kishimoto, T. Nakamura, S. Nanao, D.
Ohsawa, R. Katano and I. Isozumi: Phys. Rev. B {\bf 56} (1997) 14267.\label{expspec}
\bibitem{JF84}
        H. J. F. Jansen and A. J. Freeman,
        Phys. Rev. B {\bf 30} (1984) 561.
\bibitem{VWN80}
        S. H. Vosko, L. Wilk  and  M. Nusair,
        Can. J. Phys. {\bf 58} (1980) 1200.
\bibitem{rf:GeKabs1}
J. Kokubun: private communication.

\bibitem{rf:GeKabs2}\label{kokubu}
J. Kokubun, M. Kanazawa, K. Ishida, and V. E. Dmitrienko: 
Phys. Rev. B {\bf 64} (2001) 073203.
\end{thebibliography}
\end{document}